\documentclass{jfm}
\usepackage{natbib,amsmath,amssymb,graphicx}
\usepackage[usenames,dvipsnames,svgnames,table]{xcolor}

\ifCUPmtlplainloaded \else
  \checkfont{eurm10}
  \iffontfound
    \IfFileExists{upmath.sty}
      {\typeout{^^JFound AMS Euler Roman fonts on the system,
                   using the 'upmath' package.^^J}%
       \usepackage{upmath}}
      {\typeout{^^JFound AMS Euler Roman fonts on the system, but you
                   dont seem to have the}%
       \typeout{'upmath' package installed. JFM.cls can take advantage
                 of these fonts,^^Jif you use 'upmath' package.^^J}%
      }
  \else
  \fi
\fi

\ifCUPmtlplainloaded \else
  \checkfont{msam10}
  \iffontfound
    \IfFileExists{amssymb.sty}
      {\typeout{^^JFound AMS Symbol fonts on the system, using the
                'amssymb' package.^^J}%
       \usepackage{amssymb}%
         
         \let\geq=\geqslant
      }{}
  \fi
\fi

\ifCUPmtlplainloaded \else
  \IfFileExists{amsbsy.sty}
    {\typeout{^^JFound the 'amsbsy' package on the system, using it.^^J}%
     \usepackage{amsbsy}}
    {\providecommand\boldsymbol[1]{\mbox{\boldmath $##1$}}}
\fi

\usepackage{overpic}
\usepackage{amsmath,mathrsfs}

\newcommand{\tens}[1]{\boldsymbol{\mathsf{#1}}}
\renewcommand{\vec}[1]{\ensuremath{\mbox{\boldmath$#1$}}}

\begin{document}

\title{Settling of an asymmetric dumbbell in a quiescent fluid}
\author{F. Candelier$^1$ and B. Mehlig$^2$}
\affiliation{$^1$University of Aix-Marseille, CNRS, IUSTI UMR 7343,
13 013 Marseille, Cedex 13, France\\
$^2$ Department of Physics, Gothenburg University,
 SE-41296 Gothenburg, Sweden}
\date{}
\maketitle
\begin{abstract}
We compute the hydrodynamic torque on a dumbbell (two spheres linked by a massless rigid rod) settling in a quiescent fluid at small but finite Reynolds number.  The spheres have the same mass densities but different sizes. When the sizes are quite different the dumbbell settles vertically, aligned with the direction of gravity, the largest sphere first. But when the size difference is sufficiently small then  its steady-state angle is determined by a competition between the size difference and the Reynolds number. When the sizes of the spheres are exactly equal 
then fluid inertia causes the dumbbell to settle in a horizontal orientation.
\end{abstract}

\begin{figure}
\begin{center}
\begin{overpic}[angle=0]{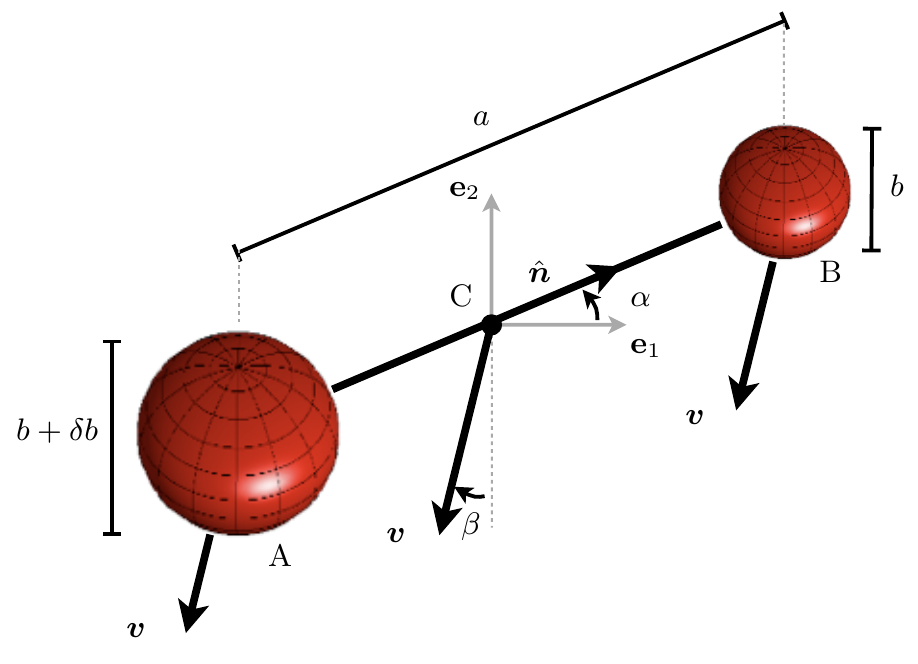} 
\end{overpic}
\end{center}
\caption{\label{fig:dumbbell}
Illustration of the dumbbell and the notation used in this
article. The centres of two spheres ($\rm A$ and $\rm B$) of diameters $b+\delta b$ and $b$ 
are linked by a massless rigid rod of length $a$. 
The centre-of-mass of the dumbbell is denoted by $C$,
and its velocity by $\vec{v}$. 
The angle of the centre-of-mass velocity
with the direction of gravity (negative ${\bf e}_2$-direction) is denoted by $\beta$.
As drawn the angle is negative, $\beta< 0$.
Since we consider a steady state where the angular velocity of the dumbbell vanishes (see text), the centre-of-mass velocities of both spheres are also $\vec{v}$.
The angle of inclination of the dumbbell (defined by the unit vector $\hat{\vec n}$) with respect to the plane orthogonal to gravity (the ${\bf e}_1$-${\bf e}_3$-plane) is denoted by $\alpha$. 
}
\end{figure}

\section{Introduction}

This paper is 
concerned with the question of how fluid inertia affects the settling of a particle under gravity in a fluid at rest.
Understanding how particles settle in a fluid is of fundamental importance in many different scientific problems. 
An important example is sedimentation in suspensions of non-spherical particles. Homogeneous 
suspensions of settling spheroids are 
unstable to the formation of
spatial patterns \citep{Koch1989,mackaplow1998,tornberg2006,metzger2007,shin2009,dahlklid2011,lundell2014}. The question is reviewed by \cite{GH11}.
Other examples are plankton dynamics in the unsteady ocean \citep{Jonsson1989,Gua12,gustavsson2016}, particle  deposition \citep{Charru2004}, and the settling of ice particles in clouds, a process relevant for the onset and nature of precipitation \citep{pruppacher2010}.

When the particle is very small, the particle dynamics can be computed in the creeping-flow limit, assuming
that inertial effects are negligible. For larger particles, when the particle Reynolds number is finite, then fluid inertia must be considered. 
It is known, for example, that weak fluid inertia causes a sphere to settle more slowly. This is a consequence
of the disturbance flow created by the sphere at small particle Reynolds number. The corresponding disturbance problem
(the \lq Oseen problem\rq{}) was solved by \cite{proudman1957} using matched asymptotic expansions. 
Fluid inertia also affects the transient settling of a sphere, before the steady state is reached \citep{Sano1981,lovalenti1993,Mei1992}.

For non-spherical particles less is known about the effects of weak fluid inertia. 
Only quite recently it was determined from first principles how weak fluid inertia modifies the torque on a spheroid in a simple shear  \citep{subramanian2005,einarsson2015a,einarsson2015b,candelier2015,rosen2015}.

How does a non-spherical particle settle in a fluid at rest?
Consider a small needle released at an inclination angle $\alpha$ with
the horizontal. If fluid inertia is negligible then the needle continues to settle at this angle if it is fore-aft symmetric. 
Its centre-of-mass moves at an angle $\beta$
with the direction of gravity (Fig.~\ref{fig:dumbbell}). This angle
is given by
\begin{equation}
\label{eq:ba_cfl}
\tan \beta= \frac{\sin(2\alpha)}{\cos(2\alpha)-3}\,.
\end{equation}
Eq.~(\ref{eq:ba_cfl}) is obtained
in the creeping-flow limit for a slender body with fore-aft symmetry,
as discussed by  \citet{Happel83}. 
See Eq.~(5-11.33) in this book.

\citet{Kha89} analysed how weak fluid inertia affects the settling needle in the slender-body approximation.
They found that a needle with fore-aft symmetry exhibits
two equilibrium orientations: vertical ($\alpha = \pi/2$) and horizontal
($\alpha = 0$). Only the horizontal equilibrium is stable. \citet{Kha89} 
concluded that the needle turns as it falls until it reaches 
horizontal orientation,  $\alpha=0$, and then it continues to fall horizontally. This implies that its centre-of-mass velocity is aligned with the direction of gravity, that is $\beta=0$.

\cite{Dabade2015} computed how weak fluid inertia affects the orientation at which spheroids with arbitrary aspect ratios settle in a quiescent fluid.
They found that when settling in a fluid at rest, weak fluid inertia causes an oblate spheroid to align its axis of 
rotational symmetry with the direction of gravity. 

Now consider a slender body that is rotationally symmetric but does not possess fore-aft symmetry. In the creeping-flow limit one expects the body to turn as it settles until it is oriented vertically ($\alpha = \pi/2$).  But fluid inertia exerts a torque that tends to turn the body towards the horizontal.
As mentioned by \cite{Kha89} this may cause the slender body to fall at an equilibrium angle   that depends upon the degree of asymmetry of the body. In this paper we calculate this angle for a dumbbell without fore-aft symmetry.
We use the method of reflection \citep{kim1991} to calculate the torque on the dumbbell, taking into account convective fluid inertia to linear order in the Reynolds number
${\rm Re}_b = b v/\nu$.  Here $b$ is the diameter of the needle, $v$ is its centre-of-mass velocity, and $\nu$ is the kinematic viscosity of the fluid.

To conclude this Introduction we briefly comment on the wider context of this work. More generally the settling of particles in turbulent flows is of interest. \cite{Maxey1987} showed that turbulence can enhance the settling speed of spherical particles. Recent experimental and numerical results on the settling of spherical particles in turbulence are discussed by \cite{Good2014}. The dynamics of neutrally buoyant  non-spherical particles in turbulence has been investigated by a number of authors \citep{Parsa2014,gustavsson2014,Byron2015,Voth2016}, but little is known about the settling of heavy non-spherical particles in turbulence.

\section{Formulation of the problem}
We consider a simple model system:
a dumbbell composed of two spheres, $\rm A$ and $\rm B$.
Their centres are linked
by a massless rigid rod of length $a$ (Fig.~\ref{fig:dumbbell}).
Both spheres have the same mass density $\rho_{\rm p}$ so that the centre-of-mass of the dumbbell coincides with its geometrical centre. 
This means that the dumbbell experiences no gravitational torque with respect to its geometrical centre. But we mention that
the effect of gravitational torques on particles with inhomogeneous mass densities (\lq gyrotaxis\rq{}) is important for the motion of certain plankton species \citep{Dur13,Bra14}. 

Sphere $\rm A$ has diameter $b+\delta b$. 
Sphere $\rm B$ has diameter $b$.
We take $\delta b>0$ and define two dimensionless  parameters that  characterise the shape of the dumbbell
\begin{equation}
\label{eq:lambda}
\lambda \equiv \frac{\delta b}{b}\geq 0 \quad \mbox{and} \quad \kappa \equiv \frac{a}{b}\,.
\end{equation}
Here $\lambda$ measures of the asymmetry of the dumbbell and $\kappa$ defines its aspect ratio.  We note
that \cite{Kha89} define their parameter $\kappa$ differently.
In the following it is  assumed that the dumbbell is slender and that its asymmetry is small:
\begin{equation}
\kappa\gg  1 \quad \mbox{and} \quad  \lambda \ll 1\:.
\label{eq_lambda_kappa}
\end{equation}
We do not consider the transient dynamics of the dumbbell but analyse 
its steady-state dynamics where the angular velocity of the dumbbell vanishes:
\begin{equation}
\boldsymbol{\omega} = \vec{0}\:.
\end{equation}
In this case the two spheres settle with the same constant velocity $\vec{v}$. We write:
\begin{equation}
\vec{v} = v \sin \beta \: {\bf e}_1 - v \cos \beta \: {\bf e}_2\:.
\end{equation}
Here $\beta$ is the angle between the velocity vector 
and the direction of gravity (the negative ${\bf e}_2$-direction, Fig.~\ref{fig:dumbbell}), and $v$ is the magnitude of this velocity. 
To completely determine the steady-state dynamics of the dumbbell it remains
to specify its angle of inclination $\alpha$ with the plane
orthogonal to gravity (Fig.~\ref{fig:dumbbell}). We write:
$$
\hat{\vec{n}} = \cos(\alpha)\, {\bf e}_1 + \sin(\alpha) \,{\bf e}_2\,.
$$
The three unknowns of the problem ($v$, $\alpha$ and $\beta$) 
are determined by three equations that are derived from the 
steady-state condition, namely that
the force on the dumbbell and the torque with respect to its centre-of-mass C  
must vanish in the steady state. The force condition reads:
\begin{eqnarray}
\label{eq_equilibrium2}
\vec{0}&=& \vec{f}_{\rm A}+ \vec{f}_{\rm B}\,.
\end{eqnarray}
Here $\vec{f}_{\rm A}$ and $\vec{f}_{\rm B}$ 
are the forces acting on spheres A and B, and $\vec{x}_{\rm A}$ and $\vec{x}_{\rm B}$ are the position vectors of their centres. The torque condition
\begin{eqnarray}
\vec{0} &=& \boldsymbol{\tau}_{\rm A}+ (\vec{x}_{\rm A}-\vec{x}_{\rm C}) \times \vec{f}_{\rm A} + \boldsymbol{\tau}_{\rm B}  + (\vec{x}_{\rm B} -\vec{x}_{\rm C}) \times \vec{f}_{\rm B} 
\label{eq_equilibrium1}
\end{eqnarray}
is a sum of the 
hydrodynamic torques centered on the spheres, $\boldsymbol{\tau}_{\rm A}$ and $\boldsymbol{\tau}_{\rm B}$,  and 
the torques with respect to
the centre-of-mass $\vec{x}_C$ of the dumbbell.
Eqs.~(\ref{eq_equilibrium2}) and (\ref{eq_equilibrium1}) 
provide three independent conditions because the forces lie in
the plane spanned by ${\bf e}_1$ and ${\bf e}_2$. The first equation gives two conditions, the second just one. 
It turns out  that the torques $\vec{\tau}_{\rm A}$ and $\vec{\tau}_{\rm B}$ are negligible (Section~3).  In this case
it follows from Eqs. (\ref{eq_equilibrium2}) and (\ref{eq_equilibrium1}) that the forces $\vec{f}_{\rm A}$ and $\vec{f}_{\rm B}$ must align with the vector $\vec{x}_{\rm A}-\vec{x}_{\rm B}$.

The spheres are subject to gravity and to 
the forces exerted by the fluid. 
The main difficulty lies in determining the force exerted by the fluid on each sphere. 
To this end we must solve the Navier-Stokes equations governing
the fluid velocity and pressure. We note that the partial time derivative of the fluid velocity evaluates to zero in the frame translating with the centre-of-mass of the particle, since the dumbbell is assumed to settle with a time-independent velocity.  In this frame of reference the Navier-Stokes equations read:
\begin{equation}
\boldsymbol{\nabla} \cdot  \vec{u} = 0\,,\quad
\rho_{\rm f} (\vec{u} \cdot\boldsymbol{\nabla} ) \vec{u}  = -\vec \nabla p + \mu \Delta \vec u + \rho_{\rm f} \: \vec g\,.
\label{eq:NS1}
\end{equation}
The boundary conditions read:
\begin{equation}
\label{eq_bdy}
\vec{u} = \vec{0} \quad\mbox{for}
 \quad \vec{x}  \in \mathscr{S}=\mathscr{S}_{\rm A} \cup \mathscr{S}_{\rm B}   \quad \quad\mbox{and}\quad \quad \vec{u} \to \vec -\vec{v}  \quad\mbox{as}\quad |\vec x | \to \infty\,.
\end{equation}
Here $\vec{u}$ is the fluid velocity as seen by the particle, $\mu = \rho_{\rm f} \nu$ is the dynamic viscosity of the fluid,
$\nu$ is the kinematic viscosity, $\rho_{\rm f}$ is its  density, $p$ is pressure and $\vec g$ is the gravitational acceleration  pointing in the $-{\bf e}_2$-direction,  and $\mathscr{S}_{\rm A}$ and  $\mathscr{S}_{\rm B}$ are the surfaces of the spheres $\rm A$ and $\rm B$.
We follow \cite{maxey1983} and decompose the fluid velocity and pressure as follows:
\begin{equation}
\vec{u} = -\vec{v} + \vec{u}^{1}  \quad \mbox{and } \quad  p = \big( \rho_{\rm f}\, \vec{g} \cdot \vec{x} + p_\infty \big) + p^{1}\,.
\end{equation}
Here  $-\vec{v}$  is the undisturbed fluid velocity in the frame translating with the centre-of-mass of the particle, $\rho_{\rm f} \,\vec{g} \cdot \vec{x} + p_\infty$ is the (undisturbed) hydrostatic pressure,
and  $\vec{u}^{1}$ and $p^{1}$ are corrections to the fluid velocity and pressure due to the disturbance caused by the dumbbell.

The forces $\vec{f}_{\rm A}$ and $\vec{f}_{\rm B}$ are given by
\begin{equation}
\vec f_{\rm A} = (m_{\rm A} - m_{\rm f A})\, \vec g + \int_{\mathscr{S}_{\rm A}} \!\!\!\boldsymbol{\sigma}^{1}\, {\rm d}\vec s\,,\quad\mbox{and}\quad 
\vec f_{\rm B} = (m_{\rm B} - m_{\rm f B})\, \vec g + \int_{\mathscr{S}_{\rm B}} \!\!\!\boldsymbol{\sigma}^{1}\, {\rm d}\vec s\,.
\label{eq_forces}
\end{equation}
The terms proportional to $\vec{g}$ are forces due to the undisturbed pressure (Archimedean forces) and gravitational forces.
Here $m_{\rm A}$ and $m_{\rm B}$  are the 
masses of the spheres and $m_{\rm fA}$ and $m_{\rm fB}$ are the 
equivalent fluid masses. The remaining terms in Eq.~(\ref{eq_forces}) are forces due to the disturbance.  
They are integrals over the fluid-stress tensor $\boldsymbol{\sigma}^{1} = -p^{1}\tens{I} + 2\mu\tens{S}^{1}$
of the disturbance.  Here  $\tens{S}^{1}$  is the symmetric part 
of the matrix of gradients of the disturbance-flow velocity $\vec u^1$. The integrals are over the surfaces $\mathscr{S}_{\rm A}$ and $\mathscr{S}_{\rm B}$ of the spheres,
${\rm d}\vec s$ is the surface element defined by the outward unit normal.  
The torques $\boldsymbol{\tau}_{\rm A}$ and $\boldsymbol{\tau}_{\rm B}$ are given by
\begin{equation}
\boldsymbol{\tau}_{\rm A} = \int_{\mathscr{S}_{\rm A}} (\vec{x} - \vec{x}_{\rm A}) \times \boldsymbol{\sigma}^{1}\, {\rm d}\vec s\,,\quad\mbox{and}\quad
\boldsymbol{\tau}_{\rm B} = \int_{\mathscr{S}_{\rm B}} (\vec{x} - \vec{x}_{\rm B}) \times \boldsymbol{\sigma}^{1}\, {\rm d}\vec s\,.
\label{eqn_torques}
\end{equation}
To determine  $\boldsymbol{\sigma}^{1}$ the disturbance problem must be solved. In the steady case the disturbance velocity $\vec u^1$ must satisfy:
\begin{equation}
\boldsymbol{\nabla} \cdot  \vec{u}^{1} = 0\,,\quad
 \rho_{\rm f}\big[ (\vec{u}^{1}-\vec{v}) \cdot\boldsymbol{\nabla} \big] \vec{u}^{1} = -\vec \nabla p^{1} + \mu \Delta \vec u^{1}\,,
\label{eq:NS1_1}
\end{equation}
with  boundary conditions 
\begin{equation}
\label{eq_bdy_2}
\vec{u}^{1} = \vec{v} \quad\mbox{for}
 \quad \vec{x}  \in \mathscr{S} \quad \quad\mbox{and}\quad \quad \vec{u}^{1}\to \vec{0}   \quad\mbox{as}\quad |\vec x | \to \infty\,.
\end{equation}
Eqs. (\ref{eq_equilibrium2}), (\ref{eq_equilibrium1}), and (\ref{eq_forces}) to  (\ref{eq_bdy_2}) constitute the problem
to be solved. 

\section{Creeping-flow limit-- method of reflection}
\label{sec:cfl}

In order to determine the forces due to the disturbance flow we must determine $\vec \sigma^1$ by solving Eq.~(\ref{eq:NS1_1}) with the boundary conditions (\ref{eq_bdy_2}). 

Consider how the orders of magnitude of the convective terms on  the left-hand side of Eq.~(\ref{eq:NS1_1}) compare to the magnitude of the viscous terms on the right-hand side of the same equation, in the vicinity of each sphere. It follows from the boundary condition Eq.~(\ref{eq_bdy_2}) that the disturbance velocity  $\vec{u}^{1}$  is of order of $v$. The length scale
is given by $b$. This allows us to estimate:
\begin{equation}
\frac{ {\rho_{\rm f}} (\vec{u}^{1} \cdot \boldsymbol{\nabla}) \vec{u}^{1} }{ \mu  \Delta \vec u^{1}} \sim \frac{{\rho_{\rm f}} (\vec{v} \cdot \boldsymbol{\nabla}) \vec{u}^{1} }{ \mu  \Delta \vec u^{1}}  \sim O({\rm Re}_b)\:.
\end{equation}
In the creeping-flow limit, ${\rm Re}_b=0$, these terms are negligible so that Eq.~(\ref{eq:NS1_1}) becomes the steady Stokes equation determining the hydrodynamic interactions between the two spheres.
We compute these interactions using the method of reflection \citep{kim1991}.
The method of reflection is applied iteratively. First the sphere $\rm A$ is considered as if this sphere was alone in a fluid at rest.   
The disturbance flow at $\vec{x}$ can then be written as:
\begin{equation}
\label{eq:uA}
\vec{u}^{1(1)}_{\rm A}(\vec{x}) = - \left(1+\frac{b^2(1+\lambda)^2}{24}\,\Delta \right) \tens{G}(\vec{x}-\vec{x}_{\rm A})\,\cdot \vec{f}^{1(1)}_{\rm A} 
\end{equation}
(the superscript in parentheses denotes the number of the iteration).
The first term is the elementary solution of Stokes equation with a point force (Stokeslet). The second term is a short-range correction due to the finite size of the particle.
Moreover
\begin{equation}
\label{eq:f11A}
\vec{f}^{1(1)}_{\rm A} = -3\pi \mu b(1+\lambda) \: \vec{v}
\end{equation}
is the Stokes drag experienced by sphere $\rm A$, $\vec x_{\rm A}$ is its position, and $\tens{G}(\vec{y})$ is  the Green tensor of the Stokes equation:
\begin{equation}
G_{ij}(\vec{y}) =\frac{1}{8\pi \mu} \Big(\frac{\delta_{ij}}{r} + \frac{y_i y_j}{r^3}\Big) \:,\quad \quad r = |\vec{y}|\:.
\label{Tensor_G}
\end{equation}
The minus sign in  Eq. (\ref{eq:uA}) arises from the fact that the sphere is seen by the fluid as a point force, with intensity equal to the Stokes drag experienced by the sphere, but with the opposite sign. 
The terms in Eq.~(\ref{eq:uA}) that account for the finite size of the particle 
are of order $(1 +\lambda)^3 b^3 v /r^3$. These terms thus contribute to the force acting on the sphere $\rm B$ at order $\kappa^{-3}$, at higher order than considered in our calculation. In the following we neglect these terms.  

Second the sphere $\rm B$ is introduced into the flow field $\vec{u}^{1(1)}_{\rm A}$, disregarding its effect on sphere $\rm A$.  To determine the force acting on sphere $\rm B$ we use the first Fax\'e{}n law \citep[see for instance][]{Happel83} which leads us to 
\begin{equation}
\vec{f}^{1(2)}_{\rm B} \simeq - 3\pi \mu b  \big[ \vec{v} -\vec{u}^{1(1)}_{\rm A}({\vec{x}_{\rm B}})  - \frac{b^2}{24} \Delta \vec{u}^{1(1)}_{\rm A}({\vec{x}_{\rm B}})\big]\,.
\end{equation}
The term involving the Laplacian is neglected in the following since
it is of order $\kappa^{-3}$. 

Third, consider the torque $\vec{\tau}_{\rm B}$. In the creeping-flow limit 
the torque centred on sphere $\rm B$ is given by 
\begin{equation}
\boldsymbol{\tau}_{\rm B}^{1(2)} \simeq \pi \mu b^3 \boldsymbol{\Omega}_{\rm f}({\vec{x}_{\rm B}}) \quad \mbox{where } \quad \boldsymbol{\Omega}_{\rm f} = \frac{1}{2} \boldsymbol{\nabla}\times \vec{u}^{1(1)}_{\rm A}\:.
\end{equation}
Since  $\boldsymbol{\Omega}_{\rm f}({\vec{x}_{\rm B}}) \sim b v/a^2$ we see that this torque is smaller than the term 
$(\vec{x}_{\rm B} - \vec{x}_{\rm C} ) \times \vec{f}^{1(2)}_{\rm B}$ by a factor of $\kappa^{-3}$.

Fourth, in the same way force and torque acting upon sphere $\rm A$ are computed. Taking the results together we find to order $\kappa^{-1}$:
\begin{eqnarray}
\vec{f}_{\rm A} &=& (m_{\rm A}-m_{{\rm f}{\rm A}})\vec{g} 
 -3 \pi\mu b(1+\lambda)\big(\vec{v} -3\pi\mu b\: \tens{G} (-\vec{a})  \cdot \vec{v} ) +  O(\kappa^{-2})\,,\\
\vec{f}_{\rm B} &=& (m_{\rm B}-m_{{\rm f}{\rm B}})\vec{g} 
 -3 \pi\mu b\big(\vec{v} -3\pi\mu b(1+\lambda)  \:\tens{G} (\vec{a})\cdot  \vec{v})+  O(\kappa^{-2})\,
. \nonumber
\end{eqnarray}
Here $\vec{a} = \vec{x}_{\rm B}-\vec{x}_{\rm A}$. Inserting these expressions into Eqs.~(\ref{eq_equilibrium2}) 
and (\ref{eq_equilibrium1}) and noting that the torques $\tau_{\rm A}$ and $\tau_{\rm B}$ are negligible
we can determine $\alpha$, $v$, and $\beta$ in the creeping-flow limit.

\subsection{Symmetric dumbbell ($\lambda=0$)}
For a symmetric dumbbell the torque condition (\ref{eq_equilibrium1}) is always satisfied in the creeping-flow limit. It follows that all inclination angles $\alpha$ are 
steady-state solutions. This well-known result \citep{Happel83} means that the dumbbell 
continues to fall at its initial inclination. 
For a symmetric dumbbell the force conditions (\ref{eq_equilibrium2}) are readily solved 
to determine the centre-of-mass speed $v$ and the angle $\beta$ as functions of $\alpha$. We find:
\begin{equation}
\label{eq:beta}
\tan \beta =  \frac{\sin(2\alpha) }{\cos(2\alpha) +3- \displaystyle{\frac{16}{3}} \kappa } 
=- \frac{ 3\sin(2 \alpha) }{16 \kappa} + O(\kappa^{-2})\,.
\end{equation}
We have expanded Eq.~(\ref{eq:beta}) for large values of $\kappa$ because our perturbative solution 
assumes that $\kappa$ is large. In this limit the angular dependence in Eq.~(\ref{eq:beta}) is qualitatively similar to that in Eq.~(\ref{eq:ba_cfl}), derived for a thin needle with fore-aft symmetry \citep{Happel83}.
According to Eq.~(\ref{eq:beta}) the angle $\beta$ is  negative for $0 < \alpha < \pi/2$ and it tends to zero as $\alpha\to 0$ and $\alpha \to \pi/2$.
An important difference between Eqs.~(\ref{eq:ba_cfl}) and (\ref{eq:beta}) is, however, that the angle $\beta$ is independent of the aspect ratio in the case of the needle, while it does depend on $\kappa$ for the dumbbell.
According to Eq.~(\ref{eq:beta}) the angle $\beta$ tends to zero in the limit $\kappa\to\infty$. This is simply a consequence of the fact that there are no hydrodynamic interactions between the spheres in this limit; they 
must fall as if they were independent, each at the same terminal velocity.

For the centre-of-mass speed we find:
\begin{equation}
\label{eq:v}
v = \frac{m_{\rm f B}  (\gamma - 1) g }{3 \pi b \mu } \left(1+\frac{6 - 3\cos(\alpha)^2}{8\kappa}\right) + O(\kappa^{-2})\,,
\end{equation}
where $\gamma = m_{\rm B}/m_{\rm f B}$. 
From (\ref{eq:v}) we infer that vertical orientation ($\alpha = \pi/2$) yields the fastest terminal velocity. We also see and that, in general, the dumbbell settles faster than either sphere A or B alone. 
This is due to the disturbance flow.

\subsection{ Asymmetric dumbbell ($\lambda \neq 0$)}
Now consider an asymmetric particle. When $\lambda$ is not zero then 
Eqs.~(\ref{eq_equilibrium2}) and (\ref{eq_equilibrium1}) admit
the solutions:
\begin{equation}
\alpha = \pm\pi/2\:, \quad \beta = 0 
\end{equation}
and 
\begin{equation}
\label{eq:v2}
 v = \frac{m_{ \rm f B}  (\gamma-1) g }{3\pi \mu b}  \Big(1+ \frac{3}{4 \kappa} +\lambda\Big) + O(\lambda^2,\kappa^{-2}, \lambda\:\kappa^{-1})\:.
\end{equation}
The asymmetric dumbbell settles vertically in the steady state. 
From the torque equation (\ref{eq_equilibrium1}) it follows that $\alpha = -\pi/2$ (the smaller sphere first) is unstable.  By contrast, $\alpha = \pi/2$ is stable.
So when an asymmetric dumbbell is released in fluid at rest then it tends to settle vertically, the larger sphere first.

\section{Effect of fluid inertia}
\label{sec:inertial}

In this Section we incorporate weak effects of fluid-inertia on the dynamics of the dumbbell settling in a quiescent fluid, 
assuming that ${\rm Re}_b$ small but finite.
As mentioned in the Introduction, \cite{Kha89} investigated the effect of fluid inertia on the settling of needle with fore-aft symmetry \citep[see also][]{Dabade2015}.  What happens when fore-aft symmetry is broken?
To answer this question we need to determine the disturbance flow produced by 
a single sphere when convective inertia is taken into account.  
The disturbance  flow produced by a single sphere  corresponds, to leading order and in the vicinity of the sphere,
 to a Stokes flow plus a small correction. Since the Stokes solution scales as $v b /r$ it follows that 
far from the spheres the convective term $\rho_{\rm f} (\vec{u}^{1} \cdot  \boldsymbol{\nabla}) \vec{u}^{1}$ is
always smaller than the viscous term $\mu \Delta \vec{u}^{1}$ and can be neglected.  But the convective term $\rho_{\rm f} (\vec{v} \cdot  \boldsymbol{\nabla}) \vec{u}^{1}$ decays more slowly than the viscous terms and must 
balance the viscous terms at a distance $b/{\rm Re}_b$, the Oseen length. At this distance this  convective term can no longer be neglected. In this way we are led 
to the Oseen equations. 

\subsection{Elementary solution of the Oseen equations}
We use an iterative procedure, as in the previous section. 
The first step is to find an elementary closed-form solution of the Oseen equations. Keeping in mind that the two spheres are assumed to be far from each other ($\kappa\gg 1$) the rapidly decreasing terms with respect to $r$ in  the solution of the Oseen equations can be neglected. 
We therefore consider the equations
\begin{equation}
\boldsymbol{\nabla} \cdot \vec{u}= 0\,,\quad
- \rho_{\rm f} (\vec{v} \cdot \boldsymbol{\nabla}) \vec{u} = - \boldsymbol{\nabla} p + \mu\Delta \vec{u}+ \vec{f} \delta\:.
\label{Green_Oseen}
\end{equation}
where the particle is represented by a  point force ($\delta(\vec{x})$ stands for the Dirac delta function).  The solution of Eqs.~(\ref{Green_Oseen}) can be found in the literature  \citep{pignatel2011}, see also (\cite{Happel83}, p. 79). 
It can be obtained as follows.
Applying the Fourier transform
\begin{equation}
\vec{\tilde{u}}= {\cal F}\vec{u} \equiv \int_{\mathbb{R}^3} \vec{u}(\vec{y}) \exp(-i\:\vec{k}\cdot \vec{y} )\:\mbox{d} \vec{y}
\end{equation}
we find
\begin{eqnarray}
\vec{k} \cdot \vec{\tilde{u}}& =& 0\,,\quad
- i \rho_{\rm f} \:(\vec{v} \cdot \vec{k}) \vec{\tilde{u}} = - i \:\vec{k} \:\tilde{p} - \mu \:k^2 \:\vec{\tilde{u}} + \vec{f} \,.\label{eq_fourier1}
\end{eqnarray}
As usual, the pressure is determined by projecting equation (\ref{eq_fourier1}) along $\vec{k}$. This yields
$ -i \:\tilde{p} = -{\vec{f}\cdot \vec{k}}/{k^2}$ and performing a partial fraction decomposition we are led to
\begin{equation}
\vec{\tilde{u}} = \frac{\vec{f}}{(\mu \vec{k}-i \rho_{\rm f} \vec{v})\cdot \vec{k}} - \left(\frac{\vec{f}\cdot \vec{k}}{\rho_{\rm f} \vec{v} \cdot \vec{k}} \right)\:i\vec{k}  \left(
\frac{1}{ k^2} - \frac{\mu}{(\mu \vec{k}-i \rho_{\rm f} \vec{v})\cdot \vec{k}}\right)\:.
\end{equation}
This result simplifies: since the vectors $\vec{f}$ and $\vec{v}$ are collinear we have 
\begin{equation}
 \left(\frac{\vec{f}\cdot \vec{k}}{\rho_{\rm f} \vec{v} \cdot \vec{k}} \right) = \frac{f}{ \rho_{\rm f} v} \:,
\end{equation}
where $f = |\vec{f}|$. 
Now we apply the inverse Fourier transform. Using the relations
\begin{equation}
\mathcal{F}^{-1} \frac{1}{ k^2}  = \frac{1}{4\pi\:r} \quad \mbox{and} \quad \mathcal{F}^{-1} \frac{1}{(\mu \vec{k}-i \rho_{\rm f} \vec{v})\cdot \vec{k}} = \frac{
\mbox{exp}\big[-\frac{1}{2\nu}( v  r +\vec{v} \cdot \vec{y})\big]}{4\pi\mu \:r}
\end{equation}
we find
\begin{equation}
\label{eq:GO1}
\vec{u} = \frac{
\mbox{exp}\big[-\frac{1}{2\nu}( v  r +\vec{v} \cdot \vec{y})\big]}{4\pi \mu \:r}\vec{f} - \frac{f}{ \rho_{\rm f} v} \boldsymbol{\nabla} \frac{1-
\mbox{exp}\big[-\frac{1}{2\nu}( v  r +\vec{v} \cdot \vec{y})\big]}{4\pi \:r}\:.
\end{equation}
Expanding the derivatives in Eq.~(\ref{eq:GO1}) yields
\begin{eqnarray}
\label{Green_Oseen_sol}\
\vec{u}  &=& \frac{
\mbox{exp}\big[-\frac{1}{2\nu}( v r + \vec{v} \cdot \vec{y} )\big]}{8\pi \mu \:r} \:\vec{f} \\
&&+ \Big\{1-
\Big(1+\frac{v\:r}{2 \nu}\Big)\mbox{exp}\Big[-\frac{1}{2\nu}(v  r + \vec{v} \cdot \vec{y} )\Big]\Big\} \,\frac{f}{\rho_{\rm f} v} \:\frac{\vec{y}}{  4\pi \:r^3}\:.
\nonumber
\end{eqnarray}
The next step is to apply the method of reflection, with Eq.~(\ref{Green_Oseen_sol}) instead of Eq.~(\ref{eq:uA}).

\subsection{Method of reflections}
As a first step  the flow $\vec{u}_{\rm A}^{1(1)}$ produced by the sphere $A$ is  considered as if it were alone in a fluid at rest. The result is given by Eq. (\ref{Green_Oseen_sol}) in which $\vec{f}$ is replaced by $3\pi b(1+\lambda) \vec{v}$  (that is the Stokes drag, with a minus sign). The sphere $\rm B$ is then introduced into this flow. We note first that the flow produced by the sphere $\rm A$  is  non-uniform. In the previous section we 
discussed the fact that  Fax\'e{}n corrections are negligible in our problem. 
One may ask whether the non-uniformity of $\vec{u}_{\rm A}^{1(1)}$ may give rise to a Saffman lift force acting on the sphere ${\rm B}$. It turns out that, in the equations governing the flow produced by the sphere $\rm B$, effects of the convective terms due to the non-uniformity of $\vec{u}_{\rm A}^{1(1)}$ are negligible   compared to those arising from the Oseen 
convective terms of the form 
$\rho_{\rm f} \big\{[\vec{v} - \vec{u}^{1(1)}_{\rm A}({\vec{x}_{\rm B}}) ]\cdot \boldsymbol{\nabla} \big\} \vec{u}_{\rm B}^{1(2)}$.  The lift force can be therefore be neglected and the equations which govern the fluid velocity  take the same form as Eqs. (\ref{Green_Oseen}) but where the velocity $\vec{v}$ has to be replaced by the relative velocity $\vec{v} - \vec{u}^{1(1)}_{\rm A}({\vec{x}_{\rm B}})$. 
The force acting on the sphere $\rm B$ \citep[see for example][]{proudman1957}  is then given by
\begin{equation}
\label{eq:fAO1}
\vec{f}^{1(2)}_{\rm B} = - 3 \pi \mu  b \big( 1 + \frac{3}{16} {\rm Re}_b \big) \big[\vec{v} - \vec{u}^{1(1)}_{\rm A}({\vec{x}_{\rm B}})\big] + O\left(\kappa^{-2}, \lambda \: {\rm Re}_b, \:{{\rm Re}_b \kappa^{-1}}\right)
\end{equation}
(note that $\vec{u}^{1(1)}_{\rm A}$ depends on ${\rm Re}_b$). Similarly, the disturbance force acting on the sphere $\rm A$ reads as
\begin{equation}
\label{eq:fAO2}
\vec{f}^{1(2)}_{\rm A} = - 3 \pi \mu b(1 + \lambda) \big( 1 + \frac{3}{16} {\rm Re}_b \big) \big[\vec{v} - \vec{u}^{1(1)}_{\rm B}({\vec{x}_{\rm A}})\big] +  O\left(\kappa^{-2}, \lambda \: {\rm Re}_b,\:{{\rm Re}_b \kappa^{-1}}\right)\:.
\end{equation}

\subsection{Results}
Inserting the disturbance forces (\ref{eq:fAO1}), (\ref{eq:fAO2}) into Eqs.~(\ref{eq_forces}) and solving Eqs. (\ref{eq_equilibrium2}) and  (\ref{eq_equilibrium1}) 
yields the desired solution. In general this must be done numerically. Here we consider the limit $\kappa{\rm Re}_b\ll1 $. In this case we can obtain explicit expressions by expanding the Oseen solution (\ref{Green_Oseen_sol}). In this way we find that the angle of inclination tends to the equilibrium value
\begin{equation}
\label{eq:result}
\alpha = \left\{
\begin{array}{ll}
\displaystyle \arcsin\Big(\frac{16\, \lambda}{3\, {\rm Re}_b}\Big)   \! +\!  { O(\kappa\:\lambda)}
 & \mbox{ when $\displaystyle \frac{16\, \lambda}{3\, {\rm Re}_b}   < \! 1$\,,}\\\displaystyle
\frac{\pi}{2} &\mbox{ otherwise\,.}
\end{array}\right.
\end{equation}
As expected it follows from Eq.~(\ref{eq:result}) that a sufficiently asymmetric dumbbell falls 
in vertical orientation. A perfectly symmetric dumbbell, by contrast, settles in the horizontal orientation, like a fore-aft symmetric slender body \citep{Kha89}.
This is a consequence of the fact that the Oseen solution breaks the symmetry of the Stokes solution.
But when the effect of breaking fore-aft symmetry and of convective fluid inertia
balance then the dumbbell falls at the equilibrium inclination determined by Eq.~(\ref{eq:result}).
For ${\rm Re}_b = 0.1$, for example
a slight asymmetry of one percent, $\lambda = 0.01$, causes the dumbbell to settle at an inclination angle different from zero.  The vector $\hat{\vec n}$ forms an angle of
roughly $30$ degrees with the horizontal according to Eq.~(\ref{eq:result}), as schematically shown in Fig.~\ref{fig:dumbbell}. 
One expects slender particles to exhibit a qualitatively similar behaviour and this could be tested experimentally 
by observing a small rod settling in a quiescent fluid 
at small Reynolds number. 

Eq.~(\ref{eq:result}) shows that the angle is independent of $\kappa$ when $\kappa {\rm Re}_b$ small. 
This is a consequence of a property of the Oseen solution
that is particular to the dumbbell. Expanding the Oseen solution assuming that $a$ is of the order of or smaller than $b/{\rm Re}_b$, we deduce that 
there is a uniform contribution to the fluid velocity that is proportional to ${\rm Re}_b$. This means that the inertial torque is proportional to $a$. Since the contribution to the torque from the particle
asymmetry is also proportional to $a$ it follows that $\alpha$ is independent of $\kappa$.  A slender body behaves slightly differently, since
the torque has a different dependence upon $a$. See 
Eqs.~(6.12) and (6.22) in \citep{Kha89}. When $\kappa {\rm Re}_b\ll 1$ the angular dependence of the inertial
torque is the same for the dumbbell as for a thin needle, proportional to $\sin(2\alpha)$.

At the order considered here, the angle $\beta$ of the centre-of-mass velocity is found to remain unchanged compared 
with the creeping-flow limit (\ref{eq:beta}).
The settling speed is given by
\begin{equation}
\label{eq:v4}
v=  \frac{m_{\rm f B}  (\gamma - 1) g }{3 \pi b \mu }  \left[\left(1+\frac{6 - 3\cos(\alpha)^2}{8\kappa}\right) + \lambda - \frac{3}{8}{\rm Re}_b \right]  + O(\epsilon^{2})\:.
\end{equation}
Here $\epsilon^2$ denotes 
any quadratic combination of the small parameters $\kappa^{-1}$, $\lambda$, or ${\rm Re}_b$.
Eq.~(\ref{eq:v4}) is consistent with the results obtained in the creeping-flow limit in Section \ref{sec:cfl}. 
In the limit ${\rm Re}_b\to 0$ Eq.~(\ref{eq:v4}) reduces to Eq.~(\ref{eq:v}) for $\alpha\neq 0$, $\lambda=0$, and to Eq.~(\ref{eq:v2}) for $\alpha=\pi/2$, $\lambda\neq 0$.
\section{Conclusions}
We computed the hydrodynamic torque on an asymmetric dumbbell settling in a quiescent fluid,
assuming that the Reynolds number ${\rm Re}_b$ is small but finite. The two spheres have the same mass densities but different sizes. This asymmetry gives rise to an additional contribution to the torque
that may balance the contribution of convective fluid inertia. In this case the dumbbell settles at an equilibrium angle that is determined by this balance. 
This prediction is expected to qualitatively hold more generally for particles of other shapes with broken fore-aft symmetry. 
When $\kappa {\rm Re}_b$ is small the equilibrium angle of the settling dumbbell is independent of the value of the aspect ratio $\kappa$. This property is particular to the dumbbell, it does not hold for more general bodies.
In an independent study, \cite{Koch2016} have analysed the settling of
small rod-like and ramified particles without fore-aft symmetry that settle in a quiescent fluid, using slender-body approximations and experiments. 
It will be of interest for future work to determine the full range of particle shapes and flow parameters at which fore-aft asymmetry produces an equilibrium sedimentation angle that is neither horizontal nor vertical.  It is of particular interest to answer the question:  how robust are results derived assuming fore-aft symmetry?

Our results were obtained using the method of reflection. This method generates a perturbation expansion in the inverse of the aspect ratio $\kappa$ of the dumbbell.
The method  makes it possible to systematically treat the dynamics of more general assemblies of spheres, partially linked. It allows, for instance, to compute the dynamics of two dumbbells settling together, affecting each other but not coming too close to each other.
The method used here is not restricted to quiescent flows. Applying the method of reflection for a neutrally buoyant symmetric dumbbell in a simple shear gives Jeffery's equation \citep{Jeffery1922}, and it may be possible to treat the effect of fluid inertia upon the angular dynamics of the dumbbell using the method of Section \ref{sec:inertial} and the reciprocal theorem \citep{subramanian2005,einarsson2015b,einarsson2015a}.

{{\em Acknowledgements}. This work was supported by grants from Vetenskapsr\aa{}det [grant number 2013-3992], Formas [grant number 2014-585], and by the grant {\em Bottlenecks for particle growth in turbulent aerosols} from the Knut and Alice Wallenberg Foundation, Dnr. KAW 2014.0048.}

\end{document}